\documentclass[prl,twocolumn]{revtex4}
\usepackage{graphicx,amssymb}
\usepackage{graphicx}
\usepackage{rotating}
\usepackage{dcolumn}
\usepackage{amsfonts}
\usepackage{amssymb}
\usepackage{amsmath}
\usepackage{latexsym}
\usepackage{epsfig}
\usepackage{dsfont,eucal,mathrsfs}
\usepackage{bm}

\usepackage{color}

\renewcommand{\mathbf}{\boldsymbol}
\renewcommand{\mathcal}{\mathscr}

\begin{document}

\title{Exponential energy growth in adiabatically changing Hamiltonian Systems}
\author{Tiago Pereira}
\author{Dmitry Turaev}
\affiliation{Department of Mathematics, Imperial College London, UK}

\begin{abstract}
Fermi acceleration is the process of energy transfer from massive objects in slow motion to light objects that move  fast.  The model for such process is a time-dependent Hamiltonian system. As the parameters of the  system change with time, the energy is no longer conserved, which makes the acceleration possible. One of the main problems is how to generate a sustained and robust energy growth.  We show that the non-ergodicity of any chaotic Hamiltonian system must universally lead to the  exponential growth of energy at a slow periodic variation of parameters.  We build a model for this process in terms of a Geometric Brownian Motion with a positive drift, and relate it to the entropy increase.
\end{abstract}

\maketitle

In his celebrated work \cite{Fermi} Fermi proposed a mechanism for particles in cosmic rays to achieve anomalously high energies. His idea can be put in a more general
framework: a fast particle can accelerate due to collisions with massive, slowly moving objects \cite{Ulam}.  This acceleration mechanism was initially motivated by plasma confinement \ \cite{plasma} and nuclear fission \cite{nuc}.
Most of the research of this phenomenon has
focused on the energy growth of particles in billiards with a moving boundary, where the bulk of numerical studies
shows the growth of the particle's kinetic energy which is at most polynomial in time \cite{Billiard,GRST}.
This regime can be easily destroyed by a small dissipation \cite{BunimovichLeonel}. However, in recent examples
a robust exponential energy growth was achieved by breaking the ergodicity of the billiard motion \cite{GRST,Batistic}.

In this Letter we investigate a general question: How a slow (adiabatic) periodic variation of parameters of an arbitrary Hamiltonian system can lead to a sustained energy growth? A theory
proposed in Refs. \cite{GRST} shows that for a billiard with slowly moving boundaries the obstacle for a fast energy growth is the ergodicity of the particle dynamics in the static billiard, which creates the so-called Anosov-Kasuga adiabatic invariant \cite{Anosov}.
This adiabatic invariant is not billiard specific, so it imposes restrictions to the energy growth in any Hamiltonian system with slowly changing parameters if the frozen dynamics is ergodic on every energy level. However, apart from special classes of systems, such as geodesic flows and billiards, a typical Hamiltonian system is not ergodic.

We demonstrate that the non-ergodicity of a chaotic Hamiltonian system must universally lead to the exponential growth of energy at a slow periodic oscillation of parameters. The key mechanism is the following: A non-ergodic Hamiltonian system has regions of chaotic dynamics in the phase space, which coexist with stability islands where dynamics is nearly integrable (quasiperiodic). Adiabatic changes of parameters lead to transitions between these regions. Different initial conditions give rise to different itineraries of these transitions, and different itineraries give different values of the energy gain/loss per period of the parameters oscillation. We show that on average over all possible itineraries the entropy of the system linearly increases after each period, which yields the exponential energy growth.

Consider a family of Hamiltonians $H(p,q;\tau)$ and assume that the parameter $\tau$ changes periodically with time. Assume the Hamiltonians in the family are homo\-geneous,
i.e., invariant with respect to energy scaling, so the dynamics in each energy level is the same. A typical example is the motion in a homogeneous polynomial potential,
see e.g. Eq. (\ref{hv4}). Another example is the Boltzmann gas of hard spheres. We focus on the homogeneous case because we are studying the process of an unbounded energy growth, and even for a general potential only the highest order terms are relevant at high energies.

We will further suppose that the system defined by the Hamiltonian $H(p,q;\tau)$ has, at each frozen value of $\tau$ from a certain region, more than one ergodic component (on each energy level). On the other hand, at $\tau$ close to the beginning of the period of the parameter oscillations we assume a strongly chaotic regime, where the dynamics mixes rapidly.

As $\tau$ changes, the energy $E=H(p,q,\tau)$ is no longer preserved by the system: $\dot{E}=(\partial H/\partial\tau) \dot\tau$. By the homogeneity, it follows that $\partial H/\partial \tau$ has the same order as $H$ at large $E$, so the speed of change of $\ln E$ is comparable with $\dot\tau$. We assume that parameters of the system change adiabatically, i.e. the change in $\tau$ and $\ln E$ is much slower than the dynamics in the $(p,q)$ phase space.

Our first claim is that at high energies we can model the energy changes by the multiplicative random walk
\begin{equation}\label{claim1}
E_{n+1} = E_n \zeta_n,
\end{equation}
where $E_n = H({p},{q}, \tau(nT))$ is the energy after $n$ periods $T$; the energy gains $\zeta_n$ form a sequence of independent, identically distributed random variables. The multiplicative character of law (\ref{claim1}) is due to $\partial H/\partial\tau \sim H$, while the randomness and independence of $\zeta_n$ is due to the chaotic behavior and fast  decay of correlations at least at a part of the period.

Model (\ref{claim1}) describes a random walk for $\ln E_n$.
It follows that the distribution of $\ln E_n$ tends to a Gaussian with the mean $n\rho$, where
$
\rho=\mathbb{E} \ln \zeta_n.
$
In particular, for a typical realization of the random walk, $\lim\frac{1}{n}\ln E_n=\rho$, so that the energy $E_n$ changes exponentially at the rate $\rho$. Note that $\mathbb{E} E_{n+1}= (\mathbb{E} \zeta_n) \mathbb{E} E_n$, so the expected value of the energy changes at a faster rate
$
\rho^+=\ln \mathbb{E} \zeta_n,$
which means that a small minority of realizations far outperform the rest. Note that similar multiplicative processes provide a basic model for describing non-thermal behavior in various applications  \cite{Ole}.

The second claim is that for adiabatically perturbed Hamiltonian systems, which have several distinct ergodic components
in the energy level, the multiplicative random walk model (\ref{claim1}) has a positive bias:
\begin{equation}\label{claim2}
\rho=\mathbb{E} \ln \zeta_n > 0.
\end{equation}
Hence,  {\it the energy grows exponentially both for typical initial conditions and on average}. Note that the non-ergodicity plays an important role here. In the ergodic case the bias $\rho$ vanishes and model (\ref{claim1}) becomes invalid (the energy grows at most polynomially in this case).

Our third claim is that in a typical situation the distribution $\nu$ of $\ln E$ is close to a Gaussian already after the first period of parameters oscillation,
so, for all $n$,
\begin{equation}\label{distLogE}
\nu (\ln E_n) \approx \mathcal N (n \rho, \sqrt{n}\sigma)
\end{equation}
where $\sigma^2=\mathbb{E} (\ln \zeta_n- \rho)^2 = 2(\rho^+-\rho)$. In other words, the energy growth is modeled by a particular class of multiplicative random processes, the so-called geometric Brownian motion (GBM).

We start with a detailed numerical verification of the above claims for an example of a particle in a quartic potential (\ref{hv4}). Then, we develop an averaging theory for non-ergodic Hamiltonians, which, in particular, implies law (\ref{claim2}). The numerical experiments are performed with
\begin{equation}\label{hv4}
H(\bm{p},\bm{q},\tau(t)) = \frac{p_1^2}{2}  + \frac{p_2^2}{2} + \frac{a(t)}{4}
\left( q_1^4 + q_2^4 \right)  + \frac{b(t)}{2} q_1^2 q_2^2,
\end{equation}
where $\bm{q} = (q_1,q_2)$, $\bm{p}=(p_1,p_2)$, and $\tau(t) = (a(t), b(t))$. For frozen values of the parameters this system has been thoroughly
studied \cite{Canergie}. For example, for $a=0.01$, $b=1$ the system exhibits exponential decay of correlations.
If $b=0$, the two degrees of freedom are uncoupled and the system is integrable. Thus,
we can change $a$ and $b$ in such a way that the system will undergo a transition between chaotic and integrable regimes,
see Fig. \ref{Fig1}a.

Numerical integration is performed using an explicit fourth-order symplectic method \cite{Sym} with integration step $h=10^{-4}$.
We always start with initial conditions uniformly distributed at the energy level $E_0 = 3 \times 10^5$.
Throughout the paper $\langle \cdot \rangle$ stands for the ensemble average with respect to these initial conditions.
The size of the ensemble is $N = 2 \times 10^4$.

\begin{figure}[!ht]
\centerline{\hbox{\psfig{file=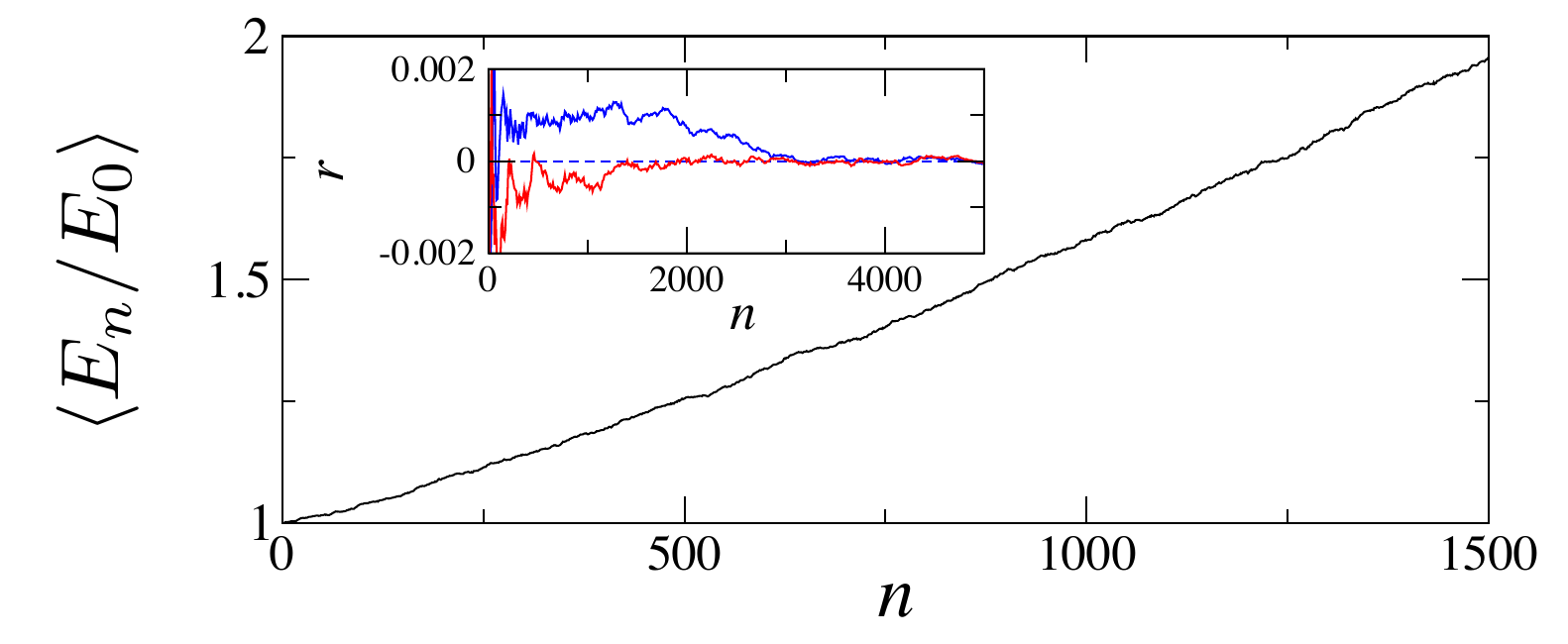,width=7.5cm}}}
\vspace{-0.5cm}
\caption{{\bf Polynomial energy growth in ergodic regime}.
We change parameters so that the frozen system remains chaotic, with no visible stability islands. The ensemble averaged energy versus time is shown.
In the inset the rates $r(n)=\frac{1}{n} \ln \left[E(n)/E(0)\right]$ are shown for two trajectories for a larger number of periods.
Both rates tend to zero, corroborating the lack of exponential acceleration.}
\label{Fig2}
\end{figure}

\begin{figure*}[!ht]
\centerline{\hbox{\psfig{file=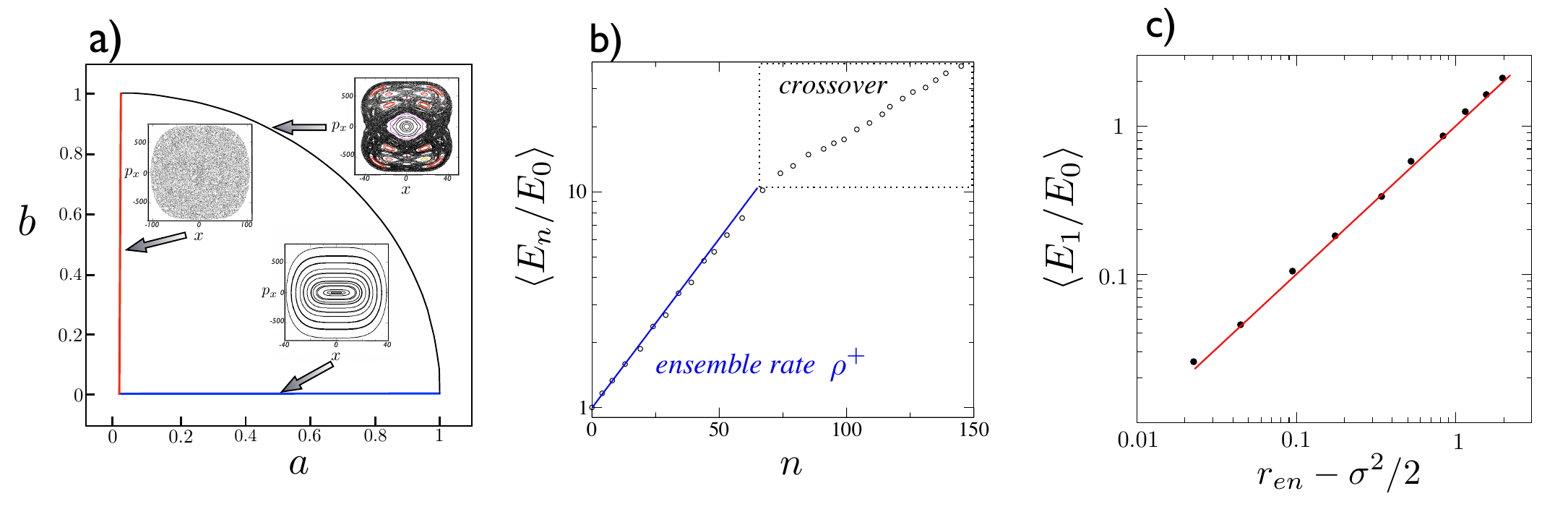,height=4.5cm}}}
\caption{{\bf Exponential energy growth}. a) Parameter space for the quartic Hamiltonian system (\ref{hv4}).
Parameters are chosen such that the frozen Hamiltonian exhibits chaotic dynamics (along $a=0.01$), quasiperiodic motion (along $b=0$), and
mixed behavior (along the connecting arc). The insets show the dynamics of the frozen Hamiltonian. Performing the cycle leads to the exponential growth of energy.
In b) we show the ensemble energy growth with the energy given in log-scale. For the first $70$ cycles the higher ensemble rate $r_{en} = 0.0368$ holds.
Further, a crossover to a lower rate starts. In c) a prediction of the
GBM model for the relation between $\rho^+$, $\rho$ and $\sigma$ is verified (the rates are varied by changing the
value of the parameter $a_0$ at fixed $A=1$). The red solid line is the identity.}
\label{Fig1}
\end{figure*}

{\it Strong chaos and polynomial energy growth.}
If we change parameters in such a way that for each frozen value of the parameters the Hamiltonian remains strongly chaotic, we observe only
a slow energy growth. For instance, Figure \ref{Fig2} shows
that for $a= 0.01$  and  $b(t) = 1.5 +\cos ( 2\pi t /T )$ with $T=400$
the ensemble energy growth $\langle E_n/ E_0 \rangle$ vs. the number of periods $n$ behaves as a quadratic polynomial.

\begin{figure}[!ht]
\centerline{\hbox{\psfig{file=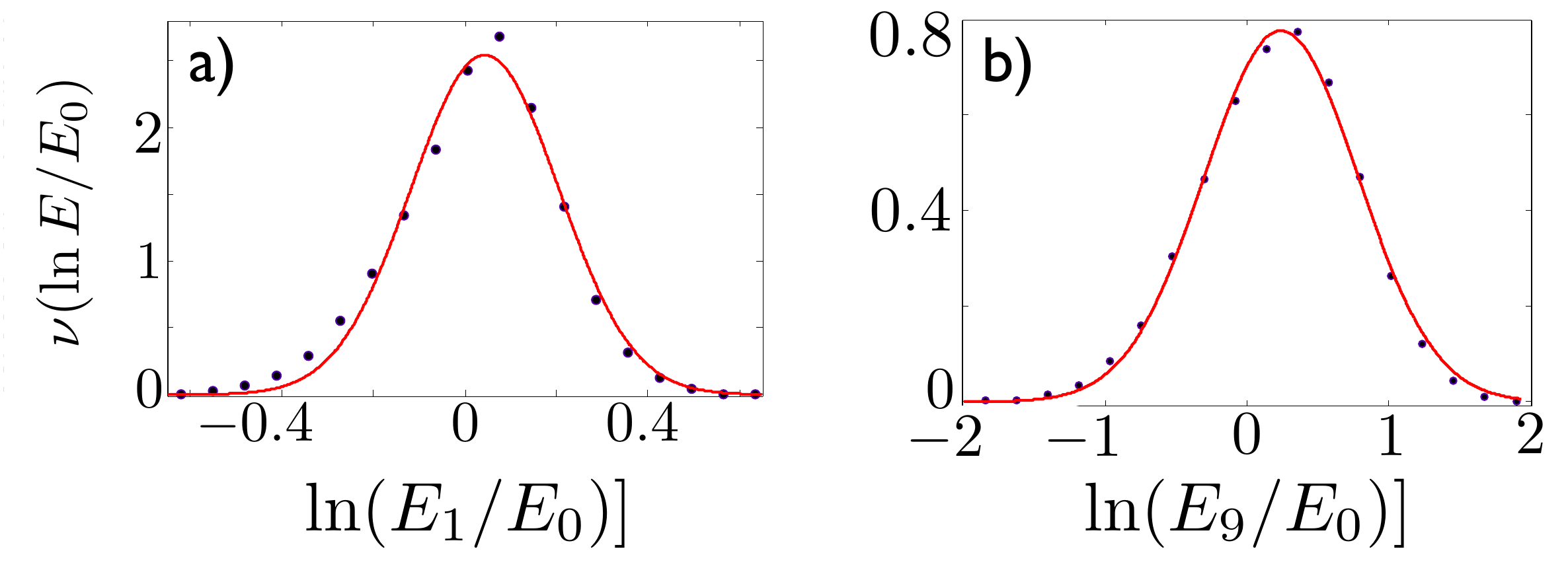,width=8.0cm}}}
\caption{{\bf Distribution of energies is log-normal}. The distribution of energies for an ensemble of $2 \times 10^4$
particles starting with initial conditions randomly distributed in the energy shell $E_0$. Already after one period
the distribution of the logarithm of $E_n$ is close to the Gauss law.}
\label{Fig4}
\end{figure}

{\it Ergodicity breaking leads to exponential acceleration.} The most of our numerical experiments correspond to the case where the parameters go through chaotic
and integrable regions in the parameter space, along the cycle displayed in Fig. \ref{Fig1}a. The cycle is described by
$a(t) = A \cos ( 2\pi t/ T)$ if $A \cos( 2\pi t /  T ) >a_0$ and $a(t) =  a_0$  otherwise,
along with $b(t) =  A \sin ( 2\pi t/ T) $  if $ A\sin( 2\pi t/T) > 0$ and $b = 0$  otherwise.
In Figs. \ref{Fig1}b and \ref{Fig4}, we show results for $a_0=0.1$, $A =1$, and $T=400$.

For each initial condition, we record the energy gain after $n$ periods: $E_n/E_0 = \Pi_{k=1}^{n-1} E_{k}/ E_{k-1}$, and compute the ensemble rate
$\displaystyle r_{en}(n) = \frac{1}{n} \sum_{k=1}^n \ln  \frac{\langle E_k \rangle}{\langle E_{k-1} \rangle}$.
The characteristic signature of GBM is that two distinct growth rates are observed when the averaging is performed over a finite ensemble.
This phenomenon is well known \cite{Ole,GRST}: As the standard deviation of $E_n$ in Eq. (\ref{claim1}) grows much faster than $\mathbb{E}(E_n)$,
it follows that for finite ensembles there is a crossover from the ensemble rate $\rho^{+}=\ln \mathbb{E} \zeta$ to a lower rate $\rho=\mathbb{E} \ln \zeta$
as $n$ grows. In Fig. \ref{Fig1}b we clearly observe this crossover to a lower (still positive) rate of the exponential energy growth. Thus,
the ensemble rate $r_{en}(n)$ observed at the initial stage of the acceleration process can be identified with the parameter $\rho^+$ of the GBM.
The data shown in Fig. \ref{Fig1}b give $r_{en}(n)$ that quickly stabilizes to $\rho^+\approx 0.0368$ and holds over the first $70$ cycles.

In order to make a qualitative check of our GBM model, we investigate the behavior of the distribution of $\ln E_n$.
As seen in Fig. \ref{Fig4}, this distribution is indeed close to Gaussian, in accordance with Eq. (\ref{distLogE}). The values of $\rho$
and the standard deviation $\sigma$ are estimated from the numerics as $\rho\approx \frac{1}{n} \langle \ln (E_n/E_0) \rangle$ and
$\sigma\approx \frac{1}{\sqrt{n}} \langle \ln (E_n/E_0) - n\rho \rangle$. In our experiment the values of $\rho$ and $\sigma$
stabilize already at the first cycle, giving $\rho = 0.0212$ and $\sigma^2=0.032$.
We performed the same numerical experiments with $a_0$ and $A$ varying in $a_0 \in [10^{-6},10^2]$ and $A\in [1, 10]$. In all experiments we observed
the log-Gaussian character of the distribution of energies established after the first cycle, with parameters $\rho>0$ and $\sigma$ independent of $n$. We also
observed the constant ensemble rate $r_{en}\approx \rho^+$ at the initial stage of the acceleration process. As a test for the Gaussianity we checked the relation
$\rho = \rho^+ - \sigma^2/2$ (which is a consequence of Eq. (\ref{distLogE})). Figure \ref{Fig1}c shows the results of this test for $A=1$.
As we see, this relation holds for the entire range of values of $\rho$; the same holds true for other values of $A$.
We conclude that the observed energy growth is governed by the GBM with a positive drift.

{\bf General Setting:} A Hamiltonian $H(q,p)$ is {\em homogeneous} if for any $E>0$ there exists a coordinate transformation $\Phi$ that keeps
the system the same, sends the energy level $H=1$ to $H=E$, and has a constant Jacobian $J(E)=E^\alpha$, $\alpha>0$. We assume that the
positive energy levels are compact, so $J(E)=V(E)/V(1)$ where $V(E)$ is the volume of the $(q,p)$-space  between the energy levels $H=E$ and $H=0$.
Thus, we can label the points in the phase space $(p,q)$ by the coordinates $(x,E)$ where $E=H(p,q)$ is the energy and $x=\Phi^{-1}(p,q)$ is the projection
to the energy level $H=1$.

Consider the family of adiabatically changing homogeneous Hamiltonians $H(p,q;\tau)$.  If the frozen system is ergodic on every energy level with
respect to the Liouville measure $\mu=\delta(E-H(p,q,\tau))dpdq$, then a theorem by Anosov
is applied \cite{Anosov} that guarantees that averaging over this
measure gives a good approximation of the slow evolution of the energy for a large set of initial conditions.


By analogy, in the case where the frozen system is {\em not ergodic} we may assume that  the slow evolution of the energy is given by
\begin{equation}\label{mna}
\dot E=\int \frac{\partial H}{\partial\tau} \delta(E-H) \mu_{\tau}(dx) \; \dot\tau
\end{equation}
where $\mu_\tau$ is, at each value of $\tau$, a certain ergodic measure on the space
of fast variables. We call the $\tau$-dependent family $\mu_\tau$ in Eq. (\ref{mna}) an {\em averaging protocol}.
It can be  different for different initial conditions, though we assume that it is independent on the initial
energy $E_0$ (by the homogeneity of the frozen Hamiltonians, this assumption is natural at large $E_0$).
Thus, we split the space of initial conditions $x$ into cells $M_1, \dots, M_k$ that give rise to distinct averaging protocols.
For each cell the majority of points exhibits the same energy evolution, while for initial conditions
from different cells the values of energy gain or loss will be different.

We assume that at some value of $\tau$ -- for example, at the beginning of the period --
the frozen system is chaotic in a sufficiently strong sense. Namely, we assume that the system  relaxes to the Liouville measure
on each energy level. This means that the distribution in the $x$-space is uniform at the beginning of each period, however, the
energy acquires a non-trivial distribution $\theta(E)$ due to different averaging protocols during the cycle. Let us show that
the entropy of this distribution is a {\em non-decreasing} function of the number of cycles.

Let $E_0$ and $E_1$ be two sufficiently large values of energy. Choose a given cell $M_k$.  If the points with initial conditions
$E=E_0$, $x\in M_k$ move to the energy  level $E=\bar E_0=e^{\lambda} E_0$ after the period of $\tau$, then the points with initial conditions
$E=E_1$, $x\in M_k$  move to the level $E=\bar E_1=e^{\lambda} E_1$, by the homogeneity of Eq. (\ref{mna}). Now note that the non-averaged
system preserves volume in the $(p,q)$-space. Therefore, it follows that the volume  occupied by the points with $x\in M_k$ between
the levels $E=E_0$ and $E=E_1$ equals to the volume occupied by the points with $x\in \bar M_k$ between the levels $E=\bar E_0$ and $E=\bar E_1$,
where $\bar M_k$ denotes the image of the set $M_k$ by the flow of the (non-averaged) system after the period of $\tau$. This gives
\begin{equation}\label{vlal}
\alpha\lambda_k=\ln(v(M_k)/v(\bar M_k))
\end{equation}
where $v$ is the volume in the $x$-space (at the level $H=1$; it is convenient to normalize $v$ so that the total volume
of the $x$-space at the beginning of the cycle equals to $1$).

When the frozen systems is strongly chaotic we can define the entropy of the system as an averaged value of $\ln (V(E)/V(1))=\ln J(E)$, that is,
$$S=\alpha \int   \ln E \;\theta(E) dE dx.$$
The change of the entropy over the period of $\tau$ is $\Delta S = \sum \alpha \lambda_k v_k$, where the sum is taken over all sets $M_k$
(each corresponding to its own averaging protocol $\mu_\tau$). By Eq. (\ref{vlal}), we obtain
\begin{equation}\label{engr}
\Delta S= \sum v(M_k) \ln  \left[\frac{v(M_k)}{v(\bar M_k)} \right]
\end{equation}
As $\sum v(M_k)=\sum v(\bar M_k)=1$ ($=$ the total volume of the $x$-space), it follows that
$
\Delta S\geq 0.
$
To see this, denote $v(M_k)=v_k$, $v(\bar M_k)=\beta_k v_k$; we have $\sum \beta_k v_k=\sum v_k=1$
implying $\Delta S= - \ln \left(\prod \beta_k^{v_k}\right)\geq -\ln \left(\sum \beta_k v_k\right)=0$.
Thus, the entropy is a non-decreasing function of time, in accordance with our claim (\ref{claim2}).

If the frozen system is ergodic for each $\tau$, the Anosov-Kasuga averaging gives $\Delta S=0$,
because $\frac{d}{dt}{V}(E,\tau)=\frac{\partial V}{\partial E} \dot E + \frac{\partial V}{\partial \tau} \dot \tau=0$
when $\dot E$ is given by Eq. (\ref{mna}), see Refs. \cite{Anosov}. This means that energy stays bounded and changes
periodically with $\tau$ (to keep $V(E,\tau)$ constant).

In the general case there is no ergodicity for all $\tau$, so there is no restriction on the growth of entropy,
and one should expect $\Delta S>0$. As there is no dependence on energy in the right-hand side of Eq. (\ref{engr}), we
get the same increment in entropy over each period of $\tau$, so $S$ must grow linearly in time. This corresponds to
an exponential growth of energy, with a rate $\Delta S/(\alpha T)$ for a typical initial condition.

The distribution of the energy gain after one cycle can, in principle, be different in different settings,
see Refs. \cite{GRST,Batistic} for billiard examples. However, in these special examples the number of different averaging protocols is small.
In a typical Hamiltonian system with the mixed phase space many different elliptic islands can coexist, so the adiabatic change of parameters
can make an orbit visit many different ergodic components with essentially random itinerary. Thus, averaging over each ergodic component
(the measure $\mu_\tau$ in Eq. (\ref{mna}) at a frozen $\tau$) results in the energy multiplied by a random factor. Since the number
of transitions between different components during one cycle of parameters oscillation is large, we obtain that
the logarithmic energy gain per period tends to Gaussian law (\ref{distLogE}).

We conclude by mentioning that our results suggest a model for an adiabatically changing Hamiltonian system as a gas of non-interacting particles
(different particles correspond to different initial conditions). As there is no interaction, there is no equilibrium distribution in energies. However,
in the ergodic case we still recover the conservation of entropy. In the non-ergodic case we can think of particles as being,
at each value of the parameter $\tau$, in different states which correspond to
different ergodic measures $\mu_\tau$ over which the averaging is performed. Thus, our gas can be considered as a mixture of different phases or fractions;
the adiabatic change of parameter can lead to particles changing their sates, so the relative densities of each fraction in the gas can vary, and this naturally leads to the entropy growth.

{\bf Acknowledgment:} We thank the hospitality of the Weizmann Institute where part of this work was written. We are grateful to V.Gelfreich and V.Rom-Kedar
for useful discussions. TP was supported by Leverhulme Trust grant RPG-279 and FP7 MC Project 303180. DT was supported by grant
14-41-00044 of RSF.

\end{document}